\documentclass[aps,amsmath,twocolumn,nofootinbib,preprintnumbers]{revtex4}
\usepackage{graphics,setspace,epsfig,color}

\usepackage[vcentermath]{}
\usepackage{epsf}
\usepackage{amscd}
\usepackage{amsmath}
\usepackage{marvosym}

\usepackage{soul}
\usepackage[normalem]{ulem}
\usepackage{color}
\definecolor{purple}{rgb}{0.75,0,0.75}

\newcommand{\beq}{\begin{equation}}
\newcommand{\eeq}{\end{equation}}
\newcommand{\bea}{\begin{eqnarray}}
\newcommand{\eea}{\end{eqnarray}}
\newcommand{\nn}{\nonumber}

\newcommand{\Msun}{M_{\odot}}

\begin{document}

\title[title]{Sound velocity bound and neutron stars}
\author{Paulo Bedaque }
\affiliation{Department of Physics,
University of Maryland College Park, Maryland 20742, USA}
\author{Andrew W. Steiner}
\affiliation{Institute for Nuclear Theory, 
  University of Washington Seattle, Washington 98195, USA}
\affiliation{Department of Physics and Astronomy, University of
Tennessee, Knoxville, Tennessee 37996, USA}
\affiliation{Physics Division, Oak Ridge National Laboratory, Oak
Ridge, Tennessee 37831, USA}


\begin{abstract}
It has been conjectured that the velocity of sound in any medium is
smaller than the velocity of light in vacuum divided by $\sqrt{3}$.
Simple arguments support this bound in non-relativistic and/or weakly
coupled theories. The bound has been demonstrated in several classes
of strongly coupled theories with gravity duals and is saturated only
in conformal theories. We point out that the existence of neutron
stars with masses around two solar masses combined with the knowledge
of the equation of state of hadronic matter at ``low'' densities
is in strong tension with this bound.
\end{abstract}

\preprint{INT-PUB-14-021}

\maketitle

\subsection{Introduction}

The nature of matter at high baryon number density is
one of the outstanding open problems and nuclear and astrophysics. In
principle, the properties of matter at densities comparable to the
nuclear saturation density ($n_0\approx 0.16/\mathrm{fm}^3$) are
determined by QCD. In practice, it has been very difficult to extract
the QCD predictions for dense matter except at extremely high
densities where asymptotic freedom allows for perturbative
calculations. The structure of large nuclei provides some information
about densities around the nuclear saturation density. Above the
nuclear saturation density all known theoretical methods break down:
nuclear effective theories break down due to the high Fermi momentum
and lattice calculations are plagued by sign problems. The only
empirical evidence we have about matter at higher baryon densities
comes from the study of neutron stars which contain matter up to 5-8
times the saturation density.

General relativity connects the equation of state of dense matter with
the relation between the radius and the mass of neutron stars.
Rotation, magnetic fields, and finite temperature make only small
corrections to the mass-radius relation. Also, we assume in this Letter
that the ground state of matter at low-densities is well-described by
laboratory nuclei. Thus, the mass-radius relation is essentially
unique, and the measurement of radii and masses of several neutron
stars determines the equation of state at high energy density. For
each equation of state there is a maximum mass beyond which no stable
configuration is possible, regardless of the radius, since a more
massive star would collapse into a black hole. The higher the pressure
for a given energy density, the larger is the maximum supported
(gravitational) mass. In the last few years, two stars were observed
with a mass around two solar masses with very small error bars. One is
a millisecond pulsar in a binary system whose mass was determined
through Shapiro delay \cite{Demorest:2010bx}; the other has a white
dwarf companion whose spectroscopy allowed a precise determination of
the neutron star mass \cite{Antoniadis:2013pzd}. These two
observations currently provide the strictest empirical constraints on
the equation of state of dense matter.

One way of characterizing dense matter is through the velocity of
sound given by\footnote{We use a system of units where $\hbar=c=1$.}
$v_s^2 = dp/d\epsilon$, where $p$ is the pressure and
$\epsilon$ the energy density (including the rest mass of the
particles). Causality implies an absolute bound $v_s\le 1$ and
thermodynamic stability guarantees that $v_s^2>0$. There are reasons,
however, to expect more stringent bounds applicable to all, or at
least a large class of materials~\cite{Cherman:2009tw}.
Non-relativistic models, at least in the range of densities where they
are applicable, predict, obviously, $v_s \ll 1$. On the other extreme
we have gases composed of ultrarelativistic (massless) particles where
$v_s^2=1/3$. The inclusion of a mass for the particles lowers the
speed of sound to $v_s^2<1/3$. Interactions among the particles, if
perturbative, also lead to $v_s<1/3$. This is the case of QCD at
asymptotically high densities (or temperatures) where a weak coupling
expansion is valid. It is thus natural to speculate that the speed of
sound at intermediate densities will interpolate between these two
limits and stay at all densities below the $v_s^2=1/3$ value, at least
in asymptotically free theories like QCD. The alternative would be the
presence of one a bump in the speed of sound at intermediate densities
before its value approaches $v_s^2=1/3$ from below asymptotically,
implying the existence of maximum and a local minimum of $v_s$ as a
function of $\mu$.

There are other reasons, to believe that the $v_s^2<1/3$ bound is
valid, even in other theories besides QCD. The value $v_s^2=1/3$ is
common to all systems with conformal symmetry, of which free massless
gases are just one example. In fact, the vanishing of the trace of the
momentum-energy tensor -- the hallmark of conformal theories --
implies that the energy density $\epsilon$ and the pressure $p$ are
related by $\epsilon=3p$ and, consequently, that $v_s^2 = 1/3$, even
in the case of strongly interacting systems.

In order to find a violation of the speed of sound bound we should
then look at strongly interacting relativistic systems away from
conformality. Strongly coupled theories are difficult to be analyse
but several calculations of the speed of sound in several different
models were performed in the strong coupling limit using the AdS/CFT
correspondence. The speed of sound was computed at high temperatures
in the single scalar model~\cite{Cherman:2009tw,Hohler:2009tv}, the
Sakai-Sugimoto model~\cite{Benincasa:2006ei} (a close analogue to
QCD), the D3/D7 system~\cite{Mateos:2007vn} and the $\mathcal{N}=2^*$
gauge theory~\cite{Benincasa:2005iv} and, in all cases, the bound
$v_s^2< 1/3$ is respected. The bound was also verified in D3/D7 system
at finite baryon and isospin chemical potential.  Each of these holographic models corresponds
to a whole family of four dimensional field theories. It is unclear,
however, how broad the set of theories covered by these examples
actually is.

Some additional insight into the physical origins of the apparent
$v_s^2<1/3$ bound can be obtained writing the (baryon number) density
$n$ as $n=N(\mu) \mu^3/(6\pi^2)$. For a free ultrarelativistic
fermionic gas $N(\mu)$ is independent of $\mu$ and equal to the number
of ``degrees of freedom" of the system (different species,
polarizations, etc.). In general, $N(\mu)$ depends on $\mu$ but we
will still refer to $N(\mu)$ as the number of effective degrees of
freedom relevant at chemical potential $\mu$. Simple thermodynamics
arguments lead to the relation
$v_s^2=\frac{1}{3}(1+\frac{\mu}{3}N'(\mu)/N(\mu))^{-1}$ so, as long as
the number of effective degrees of freedom increases with $\mu$ (and
the density) the velocity bound is valid. A similar argument can be
made for the finite temperature case by substituting $\mu$ and
$n(\mu)$ by the temperature $T$ and the entropy density $s(T)$. In
finite temperature QCD, the degrees of freedom at small temperatures
are the pions and at high temperatures the much more numerous gluons
and quarks. Lattice QCD calculations show that $N(T)$ is indeed a
monotonically increasing function of $T$ \cite{Borsanyi:2010cj} and
the bound $v_s^2<1/3$ is valid. It is much less clear whether a
similar thing happens at finite chemical potential. Some
arguments~\cite{Appelquist:1999vs}, suggest that the related quantity
$\tilde N(T)=-f(T)/T^4$, where $f(T)$ is the free energy density, is
an increasing function of $T$ in asymptotically free theories, a
result similar in spirit to the ``a-theorem"~\cite{Komargodski:2011vj}
valid for all local, unitary field theories. 
   
There are counterexamples to the bound $v_s^2<1/3$. Non-relativistic
models lead to $v^2_s>1/3$, and even $v^2_s>1$, at high densities
where they are not applicable. The well-known counterexample of
Zeldovich~\cite{zeldovich} relies on semi-classical arguments, mean
field approximations and the neglect of retardation effects. Perhaps a
better counterexample is the case of QCD with an {\it isospin}
chemical potential $\mu_I$ larger than the pion mass but smaller than
QCD scales. The isospin chemical potential drives the formation of a
pion condensate (one also has to assume that electromagnetism is
``turned off" to allow for charged pion condensation) and the energy
density oscillations on top of the condensate violates the velocity
bound, as a simple chiral perturbation theory calculation
shows~\cite{Son:2000xc}. Notice that in this case the medium is
comprised of a condensate of bosons and there is no net baryon number,
a situation physically very different to the finite density of baryon
number we are interested in.

The purpose of this paper is to demonstrate that there is an acute
tension between the $v_s^2 < 1/3$ conjecture and the existence of
neutron stars with masses $M\approx 2 \Msun$ for all reasonable low
density equations of state. This tension, for two equations of state,
was already observed in \cite{PhysRevD.88.083013} (see also a related
earlier work in ref.~\cite{Nauenberg73,Lattimer10,Lattimer12}).
Assuming the validity of the sound speed bound, the properties of
strongly-interacting matter at low density are known well enough to
put a bound on the largest star mass achievable. Because the equation
of state is very constrained up to baryon number densities about $2
n_0$, the increase of the pressure with the density is limited by the
assumption $v_s^2 = dp/d\epsilon <1/3$. In this case, the equation of
state with the largest maximum mass is that with the largest pressure
above $2 n_0$~\cite{Rhoades74,Koranda97}. As a consequence, there is a
bound on the largest neutron star mass consistent with fairly well
stablished facts about the low density behavior of mass and the bound
$v_s^2<1/3$. The remainder of this paper will demonstrate that the
numerical value of this bound is near $2 \Msun$ and to quantify the
uncertainties.
  
\subsection{The equation for state for $n<2 n_0$}
  
For densities below $2 n_0$, a non-relativistic model of nucleons
interacting through a (possibly momentum-dependent) potential is
adequate. The interactions of the nucleons in the relevant energy
regime are well known experimentally and are well fit by several
potential models. Modern Monte Carlo methods are capable of using
those to determine the spectrum of light nuclei and bulk matter with
negligible numerical error. The hierarchy observed between two and
three-body forces as well as different components of the three-body
force follow the expectation of effective theory power counting
arguments (for a review see, for instance,
ref.~\cite{Bedaque:2002mn}). 

The two-body force obtained from the chiral low momentum expansion
fits the scattering data very well. Many-body calculations using the
two and three-body forces up to next-to-next-to leading order in the
low momentum expansion were argued to be perturbative in
\cite{Hebeler:2009iv} and the neutron matter equation of state
computed in \cite{Hebeler:2010jx, Hebeler:2013nza}. Similarly, the
equation of state of pure neutron matter with the AV8$^{\prime}$
two-body force (which fits all $s$ and $p-$wave phase shifts up to
energies in excess of the ones found in back-to-back scattering of
neutrons on the Fermi surface at $n=2n_0$) and a variety of
three-nucleon forces fit to reproduce the binding energy of nuclear
matter was computed in
refs.~\cite{Gandolfi:2011xu,Steiner:2012,Gandolfi:2013baa} with
numerical error smaller than $2\%$. The different three-body forces
lead to different equations of state at high densities but, up to
densities $n<2n_0$, their effect is modest. We can see in
ref.~\cite{Gandolfi:2011xu} that the difference in the energy per
neutron at $n=2n_0$ between two extreme models (no three-body force
and the strongly repulsive Urbana IX (UIX) three-body force) is about $2$ MeV
(when the three-body forces are tuned so the binding energy of nuclear
matter at saturation is fixed) to $12$ MeV (when the three-body forces
change to cover the range of empirically allowed values of nuclear
binding). This is to be compared to the total energy per neutron which
is dominated by the rest mass $M_N=939$ MeV. This approach gives, for
densities $n< 2 n_0$ very similar results, and with similar
uncertainties, to the one in refs.\cite{Hebeler:2010jx,
  Hebeler:2013nza}
  
In a real star, the weak interactions allow for the $\beta-$decay of
neutrons into protons and a small proton fraction, $x=n_P/n < 6\%$, is
expected. In order to incorporate this information into the small
extrapolation from neutron matter (with $x=0$) to $\beta-$equilibrated
matter we use the Skyrme-like parametrization
\cite{Skyrme:1959,Hebeler:2013nza} :
\bea
\label{eq:epsilon}
\frac{\epsilon(n,x)}{n} &=&(1-x) M_N + x M_P 
\nonumber \\ && +
\frac{3 T_0}{5} \left[ x^{5/3} + (1-x)^{5/3}\right]
\left( \frac{2n}{n_0}\right)^{2/3}\nn\\
&-& T_0\left[ (2\alpha-4\alpha_L)x(1-x) + 
\alpha_L \right]\frac{n}{n_0}\nn\\
&+& T_0\left[ (2\eta-4\eta_L)x(1-x) + \eta_L \right]
\left(\frac{n}{n_0}\right)^\gamma,
\eea 
with $T_0=(3\pi^2 n_0/2)^{2/3}/2M_N$. When reduced to pure neutron
matter ($x=0$), eq.~\ref{eq:epsilon} fits the 
results of refs.
\cite{Hebeler:2010jx,Gandolfi:2011xu,Steiner:2012,Gandolfi:2013baa,Hebeler:2013nza}
and very well and it is a convenient manner to parametrize them.
Choosing the parameterization of ref.~\cite{Gandolfi:2011xu} would
give similar results to those we report. 
    
The five parameters $\alpha, \alpha_L,\eta,\eta_L$ and $\gamma$ can be
determined by the empirical knowledge of five quantities:
\bea
\label{eq:empirical}
-B &=& \frac{\epsilon(n_0,1/2)}{n_0}-\frac{M_N+M_P}{2},\nn\\
p&=&n^2 \frac{\partial (\epsilon/n)}{\partial n}|_{n=n_0, x=1/2}=0,\nn\\
K &=& 9 n_0^2 \frac{\partial^2 (\epsilon/n)}{\partial n^2}|_{n=n_0, x=1/2},\nn\\
S &=& \frac{1}{8 n_0}\frac{\partial^2\epsilon}{\partial x^2}|_{n=n_0, x=1/2},\nn\\
L &=& \frac{3n_0}{8}\frac{\partial^3(\epsilon/n)}{\partial n\partial x^2}|_{n=n_0, x=1/2}.
\eea 

The analysis of nuclear masses predicts $B=16\pm 0.1$ MeV and $n_0 =
0.16\pm0.01~\mathrm{fm}^{-3}$~\cite{Kortelainen14} and the study of
giant resonances imply $K=235\pm 25$ MeV for the nuclear
incompressibility. Finally, a wide range of experimental data from
nuclear masses, dipole polarizabilities, and giant resonances implies
$S=32 \pm 2$ MeV for the symmetry energy and $L=50 \pm 15$ MeV (see
\cite{Lattimer:2012xj,Lattimer14a} and references therein) Given
values of $B$, $n_0$ and $K$, one can determine $\alpha$, $\eta$, and
$\gamma$, and then $S$ and $L$ can be used to obtain $\alpha_L$ and
$\eta_L$. After a set of parameters is chosen, the
$\beta-$equilibrated state is found by minimizing $\epsilon(n,x)$ in
relation to $x$ for any given value of $n$. At the highest density
considered and for all parameters used $x<6\%$, confirming that only a
slight extrapolation for the pure neutron case is necessary.
  
\subsection{Bound on neutron star masses}
  
We will now determine the highest neutron mass achievable assuming the
validity of the bound $v_s^2<1/3$ and the knowledge on the low density
equation of state discussed in the previous section. 
Within the set of equations of state satisfying the low density and
the $v_s^2<1/3$ constraints, the equation of state with the largest
pressure is given by
\beq
\epsilon(p) = \left\{
\begin{array}{l}
{\rm min}_x \epsilon(n(p),x), \ \ n<2n_0\\
{\rm min}_x \epsilon(2n_0,x)+3 p, \ \ n> 2 n_0 \\
\end{array}
\right.
\eeq 

\begin{figure}[t]
  \centerline{\includegraphics[width=8cm]{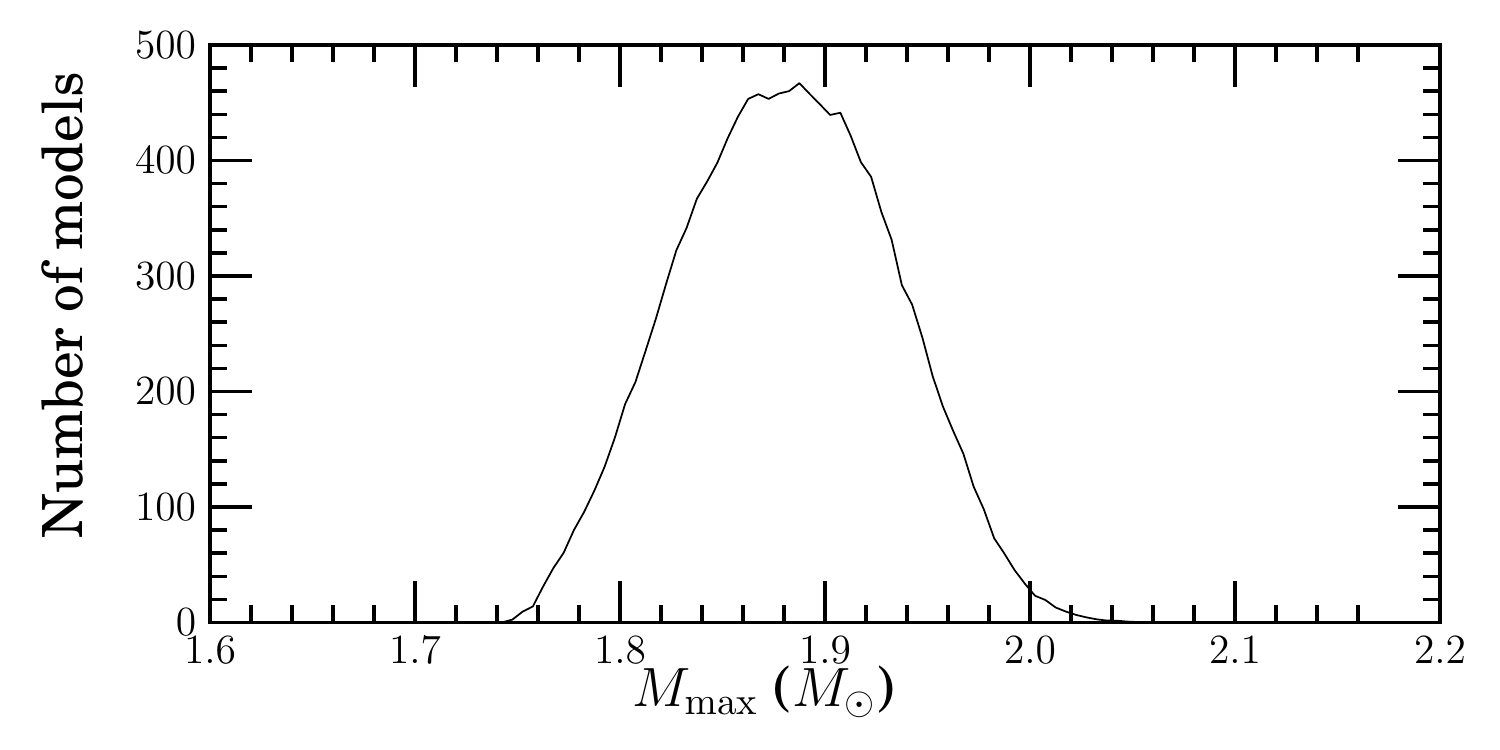}}
\noindent
\caption{Histogram of the number of models as a function of the
maximum mass supported.}
\label{fig:histogram}
\end{figure}

To reflect our uncertainty of the low density equation of state we
choose the parameters $\alpha, \alpha_L,\eta,\eta_L$ and $\gamma$ in
eq.~\ref{eq:epsilon} by selecting values for $K, S$, and $L$ at random
with a gaussian distribution centered around their empirical central
values and standard deviation given by uncertainty of their empirical
determination. Note that increasing the transition density ($ 2 n_0$)
would require massive stars to have a larger sound speed, and lowering
it signficantly would conflict microscopic calculations of the
equation of state. The small uncertainties in $B$ and $n_0$ do not
affect our results. Notice that each of these equations of state are
not meant to be realistic at high densities; they are continuous but
the speed of sound has a sudden jump at $n=2n_0$. Rather, they provide
an upper bound on the pressure for each value of the pressure and, by
the result in ref.~\cite{Rhoades74}, an upper bound on the maximum
mass of the star. For each of these equations of state (namely, for
each value of $\alpha, \alpha_L,\eta,\eta_L$ and $\gamma$) the
Tollman-Oppenheimer-Volkov equations, describing the structure of a
spherically symmetric star, is solved and the maximum mass allowed is
determined. The result is shown in the histogram in
Fig.~\ref{fig:histogram}.

The most important feature of Fig.~\ref{fig:histogram} and the main
point of this paper is the abrupt disappearance of viable models at
masses larger than about $2\Msun$. We will refrain from identifying
the number of models capable of sustaining masses above $2\Msun$ to a
probability as the error bars in the input parameters of the low
density equation of state are dominated by systematic errors. Still,
Fig.~\ref{fig:histogram} makes clear that the $v_s^2<1/3$ bound is in
strong tension with known empirical facts. This conclusion is even
more believable if one notices that we have intentionally left out
phenomena -- like the appearance of hyperons and other degrees of
freedom -- that would further decrease the pressure but that are less
certain and harder to quantify. Also, since our purpose was to
establish an upper bound on the maximum mass, we used equations of
state where the speed of sound changes suddenly from its value at
$n=2n_0$ to $v_s^2=1/3$. A smoother, more realistic transition would
further reduce the maximum mass.

\begin{figure}[t]
  \centerline{\includegraphics[width=8cm]{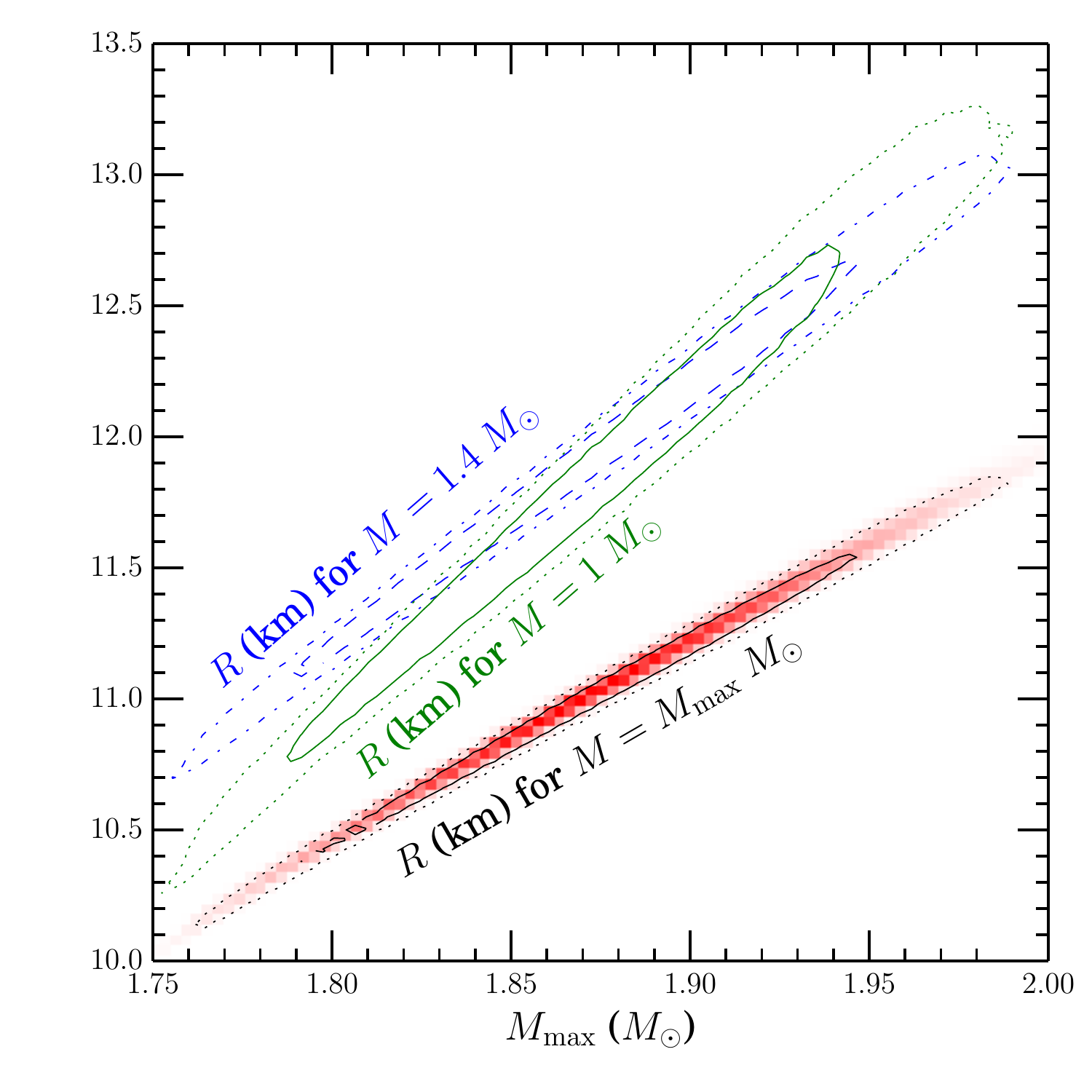}}
\caption{Radii versus maximum mass for configurations with the largest
  possible pressure subject to the velocity bound. The 68 and 95\%
  confidence regions are outlines for the radii for three different
  masses. }
\label{fig:mmax_r}
\end{figure}  

The observation of neutron stars with small radii tends to strengthen
the argument that the velocity bound must be violated. The correlation
between the radius of a 1.0 $\Msun$ neutron star and the maximum mass
is displayed in Fig.~\ref{fig:mmax_r}. The observation of a 1.0
$\Msun$ neutron star with a radius smaller than 13 km, or the
observation of any neutron star with a radius less than 11.8 km, means
that the velocity bound must be violated. In particular, the neutron
star in the globular cluster NGC 6397 already suggests that the
velocity bound must be violated, but there are several systematic
uncertainties which make this connection less
clear~\cite{SLB13,Lattimer14b}.
  
If the bound on the speed of sound is actually violated -- as it
is strongly suggested by our results-- the speed of sound, as a
function of the energy density, has a peculiar shape. It raises from
small values, reaches a maximum with $v_s^2 >1/3$, lowers to a local
minimum with $v_s^2<1/3$ and then raises again approaching $v_s^2=1/3$
from below at high densities. We find remarkable that such a
conclusion can be derived from well established facts.

There is, however, another way of looking at our result. If a proof of
the speed of sound bound is obtained, either by adapting the arguments
in refs.~\cite{Appelquist:1999vs,Komargodski:2011vj} or by other
means, our results imply that the equation of state of QCD at finite
density would be essentially known up to several times nuclear
saturation densities as only models that at low density are the
hardest allowed by empirical evidence and rapidly transition to one
with $v_s^2=1/3$ can support stars as heavy as two solar masses. Of
course, the determination of the equation of state within such a
narrow range has been a ``holy grail" of nuclear and astrophysics
since the discovery of pulsars. In addition such a result would imply
that other degrees of freedom, like $\Lambda$ hyperons cannot
appear in neutron stars in any significant numbers,
which requires strong repulsion between $\Lambda$ and
neutrons~\cite{Lonardoni14}. We would also know
that neutron stars have radii on the upper range of the current
estimates with important consequences for the detection of
gravitational waves generated in neutron star
collisions~\cite{Wade14}. The importance of all these questions
seem to warrant further field theoretical studies on the status of the
speed of sound bound. Hopefully the present paper, by pointing the
phenomenological consequences that such a proof would have, will spark
an interest in this question.

\begin{acknowledgments}
The authors would like to thank Aleksey Cherman, Tom Cohen, Aleksi
Kurkela, Shmuel Nussinov, Sanjay Reddy, and Aleksi Vuorinen for
conversations on the topic. This material is based upon work supported
by the U.S. Department of Energy Office of Science, Office of Nuclear
Physics under Award Number DE-FG02-93ER-40762.
\end{acknowledgments}

\bibliographystyle{apsrev}
\bibliography{velocity_bound.bib} 

\end{document}